\documentclass[twocolumn,dvipsnames,]{aastex63}

\graphicspath{{./}{figures/}}
\submitjournal{ApJ}

\shorttitle{Orbit Models}
\shortauthors{Pilawa et al.}

\makeatletter

\newcommand\notsotiny{\@setfontsize\notsotiny\@vipt\@viipt}
\makeatother

\newcommand{\mdm}{\ensuremath{M_{15}}}

\newcommand{\mbh}{\ensuremath{M_\mathrm{BH}}}
\newcommand{\ml}{\ensuremath{M^*/L}}

\newcommand{\GG}[1]{}
\newcommand{\tmaj}{\ensuremath{T_\text{maj}}}
\newcommand{\tmin}{\ensuremath{T_\text{min}}}
\usepackage{xcolor}
\usepackage{todonotes}

\newcommand{\cpm}[1]{{\color{black} #1}}

\let\tablenum\relax
\usepackage{siunitx}
\usepackage{amsmath}
\usepackage{multirow}

\newcommand{\vect}[1]{\boldsymbol{#1}}

\newcommand{\TriOS}{{\texttt{TriOS}}}
\newcommand{\meff}{\ensuremath{m_\textrm{eff}}}
\newcommand{\nkin}{\ensuremath{N_\textrm{kin}}}
\newcommand{\nbin}{\ensuremath{N_\textrm{bin}}}

\newcommand{\ndither}{\ensuremath{N_\textrm{dither}}}
\newcommand{\chisq}{\ensuremath{\chi^2}}

\newcommand{\xz}{\ensuremath{x\text{--}z}}

\usepackage[at]{easylist}

\begin{document}

\title{\TriOS\ Schwarzschild Orbit Modeling: Robustness of Parameter Inference for Masses and Shapes of Triaxial Galaxies with Supermassive Black Holes}

\correspondingauthor{Jacob Pilawa}\email{jacobpilawa@berkeley.edu} 

\author
[0000-0001-7040-9117]
{Jacob Pilawa}
\affiliation{Department of Astronomy, University of California, Berkeley, California 94720, USA}

\author
[0000-0002-7703-7077]
{Emily R. Liepold}
\affiliation{Department of Astronomy, University of California, Berkeley, California 94720, USA}

\author
[0000-0002-4430-102X]
{Chung-Pei Ma}
\affiliation{Department of Astronomy, University of California, Berkeley, California 94720, USA}
\affiliation{Department of Physics, University of California, Berkeley, California 94720, USA}

\begin{abstract}
Evidence for the majority of the supermassive black holes in the local universe has been obtained  dynamically from stellar motions with the Schwarzschild orbit superposition method.  However, there have been only a handful of studies using simulated data to examine the ability of this method to reliably recover known input black hole masses \mbh\ and other galaxy parameters.
Here we conduct a comprehensive assessment of the reliability of the triaxial Schwarzschild method at {\it simultaneously} determining \mbh, stellar mass-to-light ratio \ml, dark matter mass, and three intrinsic triaxial shape parameters of simulated galaxies.  For each of 25 rounds of mock observations 
using simulated stellar kinematics and the \TriOS\ code, we derive best-fitting parameters and confidence intervals after a full search in the 6D parameter space with our likelihood-based model inference scheme.
The two key mass
parameters, \mbh\ and \ml, are recovered within the 68\% confidence interval,
and other parameters
are recovered between 68\% and 95\% confidence intervals.
The spatially varying velocity anisotropy of the stellar orbits is also well recovered.
We explore whether the goodness-of-fit measure used for galaxy model selection in our pipeline is biased by variable complexity across the 6D parameter space.
In our tests, adding a penalty term to the likelihood measure either makes little difference, or worsens the recovery in some cases.

\end{abstract}

\keywords{galaxies: elliptical and lenticular, cD
--- galaxies: evolution
--- galaxies: kinematics and dynamics
--- galaxies: stellar content
--- galaxies: structure
--- dark matter}

\section{Introduction} 

Since its introduction, the orbit superposition technique of \citet{schwarzschild1979} has become a useful tool for performing detailed dynamical modeling of the internal structures of galaxies. By integrating a set of representative stellar orbits in an assumed gravitational potential, a superposition of orbits is constructed to replicate the stellar kinematics and surface brightness profiles of a galaxy seen in projection. Schwarzschild's method was initially proposed to demonstrate the existence of self-consistent galaxies with triaxial mass distributions.  It has since been extended to allow for fitting of kinematic and photometric observations (e.g., \citealt{pfenniger1984,richstonetremaine1984,richstonetremaine1985,rixetal1997}) and  is used to constrain properties of the host galaxy such as its supermassive black hole mass \mbh, stellar mass-to-light ratio \ml, dark matter content, intrinsic shape, and stellar orbital structure.

Despite the wide application of the Schwarzschild method to real data, its ability to recover known input parameters has only been tested in a handful of cases, each case with different underlying assumptions.
Early tests of the Schwarzschild method on simulated data in the three-integral axisymmetric limit highlighted the advantages and challenges posed by the method when it was applied to long-slit kinematic data (e.g., \citealt{vallurietal2004, crettonemsellem2004, magorrian2006}).
The availability of integral-field spectroscopy (IFS) over the past two decades has led to significantly improved data quality and number of kinematic constraints. An early application of axisymmetric orbit modeling to Sauron IFS data of M32 yielded well constrained \mbh, \ml, as well as the inclination angle, $i=70^\circ \pm 5^\circ$ \citep{verolmeetal2002}.
A subsequent study of NGC~2974 (also using Sauron IFS data) obtained well constrained \ml\ and $i$ from axisymmetric modeling of real data, but tests on simulated data found the inclination to be marginally constrained; the authors cautioned the validity of the formal orbit model solution $i=65^\circ\pm 2.5^\circ$ \citep{krajnovicetal2005}.
Since then, other tests on simulated data using different axisymmetric orbit codes have found a range of results. For instance, Appendix A.3 of \citet{siopisetal2009} and Appendix C of \citet{quennevilleetal2021} reported excellent recoveries of \mbh\ and
halo circular velocity;
\citet{thomasetal2007} was able to use the orbit method to reconstruct the stellar masses and velocity anisotropies in $N$-body merger remnants with high accuracy;
\citet{vasilievvalluri2020} 
recovered the true values of \ml\ but were unable to put strong constraints on \mbh\ and dark matter halos; \citet{lipkaetal2021} discussed the tendency for the inclination angle to be biased towards edge-on due to increased model flexibility.

When the axisymmetric assumption is removed, the Schwarzschild method becomes significantly more complicated and computationally expensive. Triaxial orbit modeling requires specification of three unequal axes (or equivalently, three angles) and integration of new libraries of stellar orbits, so a full exploration of the galaxy parameter space is much more computationally intensive than for axisymmetric models. Accordingly, there have been fewer attempts at performing recovery tests on simulated data.
Early tests on mock triaxial Abel galaxies
recovered \ml\ and intrinsic axis lengths  at 10\%-20\% accuracy
\citep{vandenboschetal2008, vandenboschvandeven2009}.
When the same triaxial code was applied to galaxies in hydrodynamical cosmological simulations, \citet{jinetal2019} found that depending on the assumed halo profile, the mean stellar mass is under-estimated by 13\%-24\% and the halo mass is over-estimated by 18\%-38\%.
All these tests, however, used the original 2008 code that had been shown to have 12 wrong signs in some components of the orbital velocities and various other issues \citep{quennevilleetal2022}.
\citet{Neureiter2023a} reported robust recovery of \mbh\ and \ml\ for several tested orientations of an $N$-body merger remnant.

In this paper, we conduct 
a comprehensive study of the ability of the triaxial orbit method to {\it simultaneously} recover the intrinsic shape and mass parameters of simulated triaxial galaxies containing SMBHs, stars, and dark matter. 
We have been working to reduce the computational cost of individual triaxial Schwarzschild models in the triaxial code  \texttt{TriOS} \citep{quennevilleetal2021, quennevilleetal2022}, as well as improve the efficiency in exploring the mass and intrinsic shape parameter space via non-grid-based, likelihood approach to parameter inference.  We apply this efficient parameter inference methodology to a suite of simulated triaxial galaxies, testing the ability of \texttt{TriOS} to recover known input parameters to our models while all parameters are varied simultaneously.

An additional aim of this paper is to investigate the extent to which our triaxial Schwarzschild orbit models are impacted by varying degrees of statistical complexity in the \texttt{TriOS} models. It has been a common practice while performing Schwarzschild modeling to compute a \chisq\ value associated with a model's goodness-of-fit to a set of observed constraints, and select the ``best-fit'' model that minimizes the \chisq. While often effective, this strategy potentially introduces biases due to the variable complexity of the underlying models. Intuitively, if models in some region of the parameter space have more flexibility to fit the observations, those models may have lower \chisq\ values as a simple consequence of that flexibility. 
A more prudent model selection procedure therefore should balance a combination of the fit quality and a measure of the complexity of a model. 
 
It is, however, non-trivial to quantify the complexity of a model.  The number of degrees of freedom (DOF) is often used for linear models without constraints because the number of non-redundant free parameters in a model provides a natural measure of the model complexity, or flexibility in overfitting.
While subtleties exist in the use of DOF for unconstrained linear models, the problem of how to capture complexity is even thornier for linear models with constraints or priors (and hence with reduced model flexibility), or for nonlinear models \citep{andraeetal2010}.
Different forms of complexity measure, or information criterion (IC), for the effective number of parameters in a model have been proposed, e.g., 
AIC \citep{akaike1973}, TIC \citep{Shibata1989}, BIC \citep{Schwarz1978}, NIC \citep{murata1994}, and a Bayesian measure \citep{spiegelhalteretal2002}. 
\citet{ye1998} suggests a generalized DOF that measures how sensitive the model predictions are to perturbations in the model constraints.
A model with higher complexity or flexibility is one whose predictions are more responsive to those perturbations.
Following \citet{ye1998}, \citet{lipkaetal2021} constructed an estimate of the complexity of axisymmetric Schwarzschild orbit models.
Here we examine the role of penalty terms for triaxial orbit models in simulated galaxies. 

The paper is laid out as follows. Section~2 summarizes the Schwarzschild orbit method and our procedure for constructing simulated galaxy kinematics. Section~3 outlines the main results of the paper, including a description of the parameter inference scheme used to compute our parameter estimates and corresponding confidence intervals. In Section 4, we briefly discuss estimating the statistical complexity of triaxial Schwarzschild models, and the extent to which taking into account the model flexibility changes our parameter estimation. Section 5 summarizes the results and the outlook for future triaxial stellar dynamical modeling efforts. 

\section{Dynamical Modeling and Simulated Data}

\subsection{Triaxial Schwarzschild Orbit Modeling}\label{sec:dynamical_modeling}

Throughout this work, we perform triaxial Schwarzschild orbit modeling with the \TriOS\ code \citep{quennevilleetal2021, quennevilleetal2022}, which is based on an earlier unnamed \cpm~{triaxial modeling} code \citep{vandenboschetal2008}.
In this method, a stationary gravitational potential is proposed, and a library of orbits spanning the phase space is integrated.
A set of weights are assigned to these orbits such that the linear superposition of the orbits can both reproduce the mass distribution associated with the gravitational potential and fit a set of kinematic observations.
This procedure enforces self-consistency between the mass distribution and gravitational potential without requiring specific assumptions about the form of the distribution function or velocity anisotropy.

The mass components of a galaxy in \TriOS\ can consist of a central SMBH with mass \mbh, a stellar component with mass-to-light ratio \ml, and a dark matter halo.
Most prior work using the Schwarzschild orbit method has assumed a spherical or axisymmetric potential; \TriOS\ relaxes these assumptions and allows triaxially-shaped 3D stellar mass densities.
The stellar mass density is determined by first modeling the 2D (observed) surface brightness distribution of a galaxy as a Multi-Gaussian Expansion (MGE) \citep{cappellari2002}. 
For a given set of angles $\theta$, $\phi$, and $\psi$ that relate the intrinsic (3D) and projected (2D) coordinate systems, we deproject each component of the MGE and sum them to create the 3D stellar density. 
Each MGE component (labelled by subscript $j$) is allowed to have its own 2D axis ratio, $q'_j$, which is the ratio of the lengths of its semi-minor and semi-major axes. Through the relations in the appendix of \citet{quennevilleetal2022}, the intrinsic axis ratios are determined: $p_j=b_j/a_j$ is the intrinsic middle-to-long axis ratio; $q_j=c_j/a_j$ is the intrinsic short-to-long axis ratio; and $u_j$ is the apparent-to-intrinsic long axis ratio.
When performing parameter searches and inferences below, we follow \citet{quennevilleetal2022} and use an alternative set of shape parameters, $T$, $\tmaj$, and $\tmin$.
The definitions and advantages of these parameters are articulated in Sec.~3 of \citet{quennevilleetal2022}.

For each galaxy model, the \TriOS\ code samples the phase space with a set of representative stellar orbits and integrates their trajectories. 
The orbits are initiated in two separate spaces: the \xz\ and the stationary start spaces,
where the $x-$, $y-$, and $z-$axes are directed along the intrinsic major, intermediate, and minor axes of the galaxy, respectively.
In each space,
$N_E$ values of the orbital energy are sampled.
In the \xz\ start space, orbits of a given energy are initialized on a polar grid in the \xz\ plane along $N_{I_2}$ rays uniformly spaced in angle from the $z$-axis to the $x$-axis, and at $N_{I_3}$ positions along each ray (see Fig.~4 of \citealt{quennevilleetal2022}).
Each orbit has initial velocities $v_x = v_z = 0$ and $v_y > 0$.
In addition, a set of `retrograde' orbits are constructed by simply inverting the velocities of orbits in the prograde \xz\ orbit library, and seven additional copies of each orbit are generated through the mirroring scheme described in 
\citet{quennevilleetal2022}.
In the stationary start space, orbits of a given energy are initiated with zero velocities at positions on the equipotential surface sampled on a uniform 
grid over the two spherical angles $\theta$ and $\phi$.
To enhance phase-space sampling, the code allows for ``dithering'' the initial conditions, by which the number of sample points along each of the three dimensions of the start space is increased by a factor of $\ndither$, resulting in an overall factor of $\ndither^3$ increase in the phase space sampling. The properties of bundles of $\ndither^3$ adjacent orbits are averaged and each bundle of orbits is given a single orbital weight when constructing the superposition.

Each orbit is projected onto the sky and its line of sight velocity distribution (LOSVD) within $\nbin$ spatial apertures is stored.
Its mass distribution in a 3D grid is also stored. 
We perform a non-negative least squares (NNLS) optimization to determine a set of orbit weights such that the superposition of orbits reproduces both the 
3D stellar mass distribution (obtained from de-projection of observed surface brightness distribution) and the observed LOSVD in each spatial bin.
The LOSVD is characterized by a set of Gauss-Hermite moments \citep{vanderMarelFranx1993}, and the NNLS objective function consists primarily of the \chisq\ associated with the reproduction of those observed moments (see Equation~\ref{eqn:logl}). The models contain \nkin\ kinematic constraints, \cpm{which is equal to the product of the number of spatial apertures and the number of Gauss-Hermite moments used to characterize the LOSVDs}.
The mass self consistency is enforced by adding additional \chisq-like terms to the NNLS objective function that are associated with the mismatch between the input's and model's mass in each of the 3D grid cells and within each of the 2D spatial apertures with a $1\%$ error imposed on each mass.
The phase-space distribution function for the galaxy model can be understood as being composed of this set of orbit weights.

\subsection{Generation of Simulated Galaxy Data}\label{sec:mock_galaxy_generation}

We use the Schwarzschild method to create simulated stellar kinematics for a suite of model galaxies, each of known mass components and a known intrinsic shape.  
For the simulated data to provide useful insight into actual observations, we opt to create stellar kinematics that mimic those observed in real galaxies.
We achieve this by building the simulated galaxy kinematics
using the measured kinematics of a real galaxy as a template.
Here we use the kinematics of NGC~2693 presented in \citet{pilawaetal2022}, which are typical of those known for massive elliptical galaxies in the MASSIVE survey \citep{maetal2014}.
Since these kinematics are only used as a starting template, we do not expect the main conclusions of this paper to depend \cpm{strongly} on the particular galaxy choice. 
\cpm{We note that while the velocity anisotropy profiles investigated here tend to be tangential at small radii and radial at large radii (Fig.~6 of \citealt{pilawaetal2022}; Fig.~\ref{fig:beta_recovery} here), this profile is not unique to NGC~2693. Similar profiles are found for many massive galaxies that have been studied with the orbit modeling method (e.g., Extended Data Figure~6 of \citealt{thomasetal2016}). 
A different procedure would be needed to produce a qualitatively different velocity anisotropy profile.}

Six parameters were determined for NGC~2693: 
SMBH mass \mbh, the stellar mass-to-light ratio \ml, the dark matter mass enclosed within a radius of 15 kpc \mdm, and three shape parameters ($T$, \tmaj, \tmin), or equivalently, the luminosity-averaged axis ratios $(p, q, u)$. The dark matter halo is logarithmic with a fixed scale radius $R_c =15$ kpc.
To generate simulated stellar kinematics for a given galaxy model $\textrm{G}_i$ listed in Table~1, we first
run \TriOS\ to compute the orbit libraries representing this galaxy and determine the set of orbit weights with a regularization scheme (see below) such that the model's projected kinematics best reproduce those of real observations of NGC~2693.
We then perturb this model's noise-free projected kinematics with Gaussian noise with seed $r$ and an amplitude set by the observed uncertainties on each measurements.

While determining the orbital weights of the simulated galaxies described above, we incorporate the distribution function regularization scheme outlined in Section 5.2 of \citet{vandenboschetal2008}.
This scheme penalizes the goodness-of-fit statistic when fitting the orbital weight distribution by a term proportional to the second derivative of the orbital weights with respect to the orbits' indices in their start space.  This procedure yields distribution functions that are substantially smoother
in phase space than can be obtained in models without the regularization.
The stellar masses 
in regularized models are represented by many more orbits than in non-regularized models.
For example, for the galaxy model that best reproduced the data in \citet{pilawaetal2022}, non-regularized models had $99\%$, $99.9\%$, and $99.99\%$ of the mass in 298, 365, and 397 of the 14040 orbits in the library, respectively; while regularized models have the same percentages of the mass represented by ${\sim}5100$, ${\sim}6400$, and ${\sim}7200$ orbits, respectively. 
Here we use regularization as a way to generate simulated galaxy kinematics that are represented by a wide variety of orbital weights.  In the tests below, we will assess the ability of the \TriOS\ code in recovering the input galaxy parameters without using regularization, as is commonly done when the Schwarzschild method is applied to real data to measure \mbh\ and other parameters.  

\begin{table}[ht]
\hspace{-3.5em}
\begin{tabular}{c l l l l l }
\hline
\textbf{Model} & $\textrm{G}_1$ & $\textrm{G}_2$ & $\textrm{G}_3$ & $\textrm{G}_4$ & $\textrm{G}_5$ \\ \hline
\vtop{\hbox{\hspace{0.75em}\strut $\mbh$}\hbox{\strut $[10^9 M_\odot]$}}       &  $2.1$       &  $1.7$       &  $1.0$       &  $2.4$       &  $1.3$       \\ 
\vtop{\hbox{\hspace{0.75em}\strut $\ml$}\hbox{\strut $[M_\odot / L_\odot]$}}   &  $2.23$      &  $2.31$      &  $2.43$      &  $2.25$      &  $2.28$      \\ 
\vtop{\hbox{\hspace{0.75em}\strut $\mdm$}\hbox{\strut $[10^{11} M_\odot]$}}    &  $5.5$       &  $8.9$       &  $10.2$      &  $5.5$       &  $6.0$       \\ 
$T$        																	   &  $0.39$      &  $0.47$      &  $0.46$      &  $0.49$      &  $0.41$      \\ 
$\tmaj$    																	   &  $0.04$      &  $0.17$      &  $0.26$      &  $0.25$      &  $0.10$      \\ 
$\tmin$    																	   &  $0.07$      &  $0.09$      &  $0.06$      &  $0.05$      &  $0.09$      \\ 
$u$        																	   &  $0.997$     &  $0.982$     &  $0.971$     &  $0.970$     &  $0.991$     \\ 
$p$        																	   &  $0.909$     &  $0.886$     &  $0.885$     &  $0.878$     &  $0.902$     \\ 
$q$        																	   &  $0.746$     &  $0.733$     &  $0.729$     &  $0.731$     &  $0.739$     \\ 
\end{tabular}
\caption{Mass and shape parameters for the five triaxial galaxy models tested in this work: \mbh\ is the SMBH mass, \ml\ is the stellar mass-to-light ratio, \mdm\ is the dark matter mass enclosed within a radius of 15 kpc, and $T$, \tmaj, and \tmin\ are the three shape parameters specifying the triaxial potential.  The axial ratios $u$, $p$, and $q$ are related to the shape parameters and are computed here by
taking luminosity-weighted averages over the MGE components (see text).}

\label{tab:mock_table}
\end{table}

In this work, we generate simulated kinematics for five different locations in the 6D parameter space, which are labeled $\textrm{G}_1$ through $\textrm{G}_5$ and listed in Table~\ref{tab:mock_table}.
These parameter values are chosen to span a broad range in each dimension, but narrow enough that the models can be explored with a reasonably sized collection of orbit models.
This allows us to test the ability of the orbit code to recover model parameters over a wide range of physically reasonable values.  We recall that  the axis ratios $(p, q, u)$ must obey the inequality $0\le q \le uq' \le p \le u \le 1$, where $q'=b'/a'$ is the observed axis ratio of the 2D projected galaxy shape (e.g., \citealt{quennevilleetal2022}),
and only a subspace of $(p, q, u)$ has 3D shapes that could produce the required flattening $q'$ upon projection.
The range of allowed $(p, q, u)$ is therefore typically quite narrow.

\begin{figure*}[htp!]
  \centering
  \includegraphics[width=5in]{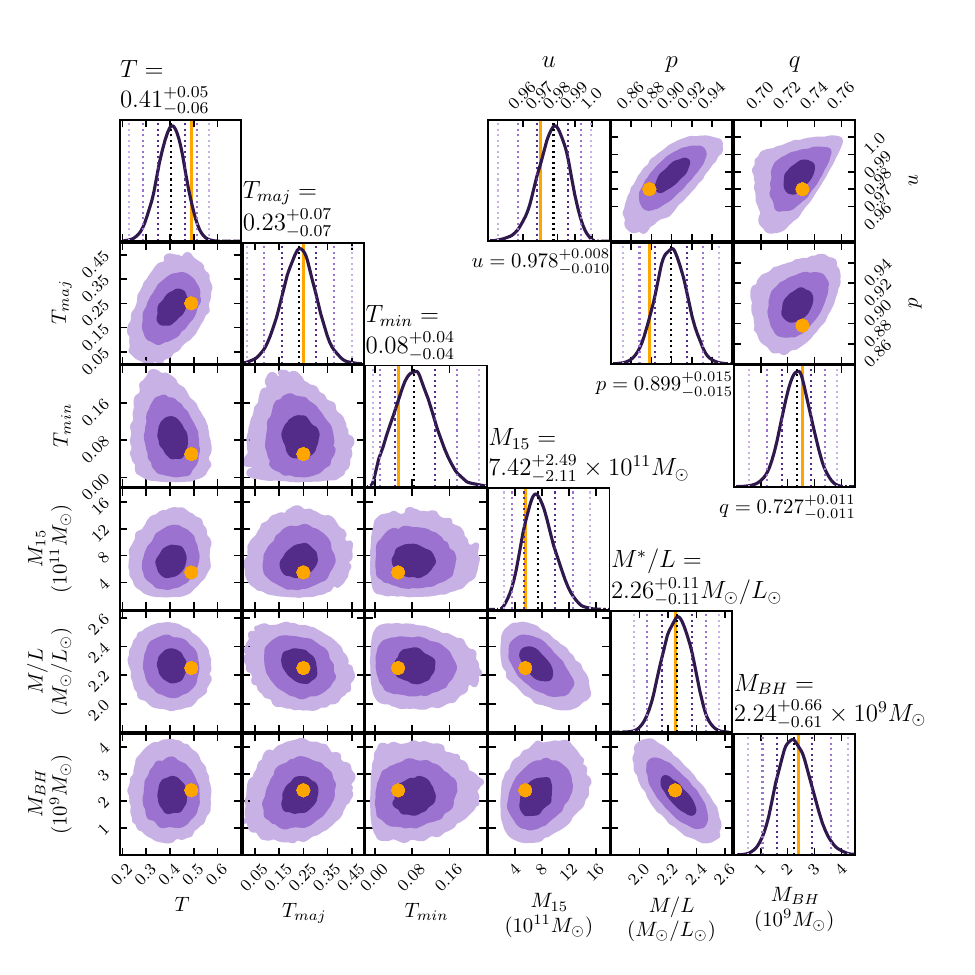}
    \caption{(Lower left) 1D and 2D marginalized posteriors of the six model parameters in an example recovery test (realization $r_5$ of model $\textrm{G}_4$).
The $68\%$, $95\%$, and $99\%$ confidence interval contours are represented by the different shades of purple in the 2D panels, and the 1D marginalized posteriors are shown in the 1D panels. The true input value for each parameter is represented by the orange filled circle and orange vertical line in each panel. (Upper right) 1D and 2D marginalized posteriors in the axis-ratio space of $(p,q, u)$, computed from the (luminosity-weighted) posteriors of $(T,\tmaj,\tmin)$. }
    \label{fig:N2693_sample_6d}
\end{figure*}

For each galaxy model $\textrm{G}_i$, we perform five draws of random Gaussian noise and 
add it to the galaxy model's noise-free projected kinematics.
The amplitudes of the injected noise are chosen to resemble those inherent in real data,
so the five realizations mimic repeated observations of the same galaxy and allow us to assess the uncertainties in the recovery test for each of the simulated galaxies.
The different noise realizations are labeled $r_1$ through $r_5$. 

In total, we have 25 sets of simulated stellar kinematics for five models of realistic galaxies.  Next, we treat each of the 25 sets of kinematics as a mock observation of a galaxy, and perform recovery tests over the full 6D parameter space to search for the best-fit parameters.

\section{Parameter Inference and Recovery Test}\label{sec:parameter_search}

\begin{figure*}[htp]
  \centering
   \includegraphics[width=6.0in]{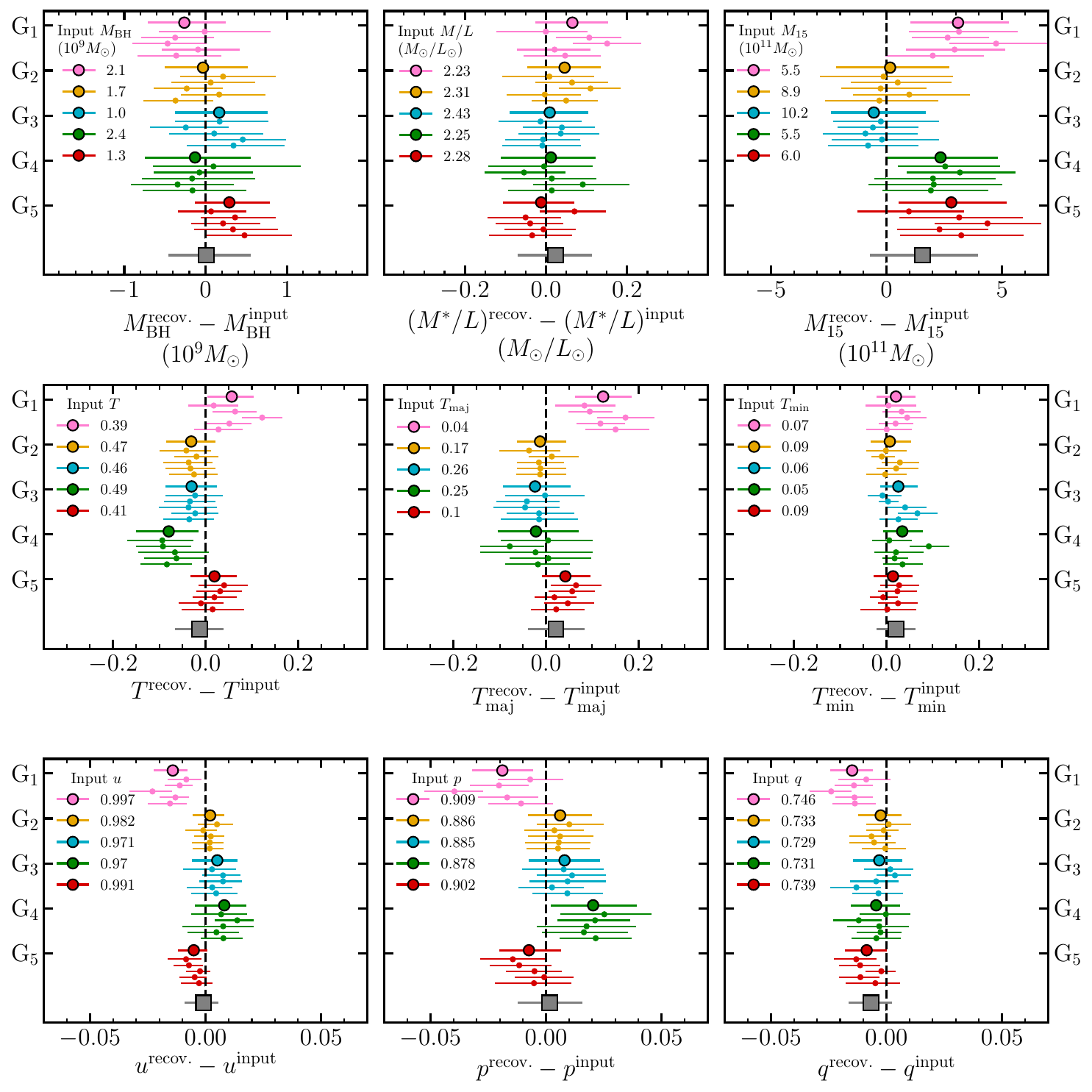}
    \caption{Results of the 25 rounds of recovery tests for the galaxy model parameters.
    The upper three panels denote the supermassive black hole mass \mbh, stellar mass-to-light ratio \ml, and enclosed dark matter mass at $15 \text{ kpc}$. The middle panels show the recovered intrinsic shape parameters $T$, \tmaj, and \tmin, and the bottom panels show the same shape recovery but for the more intuitive luminosity-averaged axis-ratios $u$, $p$, and $q$.  In each panel, the difference between the input parameter value (listed in the legend) and the recovered value is shown; the black dashed vertical line indicates exact recovery of that parameter.
    The five galaxy models $\textrm{G}_1$ through $\textrm{G}_5$ are grouped by colors.
    Within each color group, the small dot denotes the recovery result for one noise realization of model $\textrm{G}_i$, \cpm{where the error bars represent the $68\%$ credible region of the posterior samples (see Fig.~\ref{fig:N2693_sample_6d} for an example).} The larger filled circle for each $\textrm{G}_i$ model is the mean recovered value for the five realizations, where the error bars denote the median uncertainty of the realizations. \cpm{The large gray square at the bottom of each panel shows the mean over all 25 recovery tests,
    where the error bars denote the median value of the uncertainties of the 25 tests.
    }
    }
    \label{fig:N2693_Triaxial_Recovery}
\end{figure*}

In this section, we present the results of the 25 recovery tests and compare the inferred model parameters against the known input parameters of each simulated galaxy to assess how well the procedure recovers the model parameters.  The computational cost is substantial and is akin to performing triaxial Schwarzshild orbit modeling and parameter inference for 25 galaxies.

For each of the five realizations for a given galaxy model in Table~1, we treat the mock galaxy's projected kinematics as simulated data and perform a 6D parameter space search following the grid-free procedure we have developed for recent analyses of real data (e.g., \citealt{quennevilleetal2022, pilawaetal2022, liepoldetal2023}).
In this procedure, a grid-free Latin hypercube scheme is used to choose sampling points in the galaxy model parameter space (about 3000 models in this work).
For each sampled model, we run \TriOS\ and compute a goodness-of-fit metric associated with the model predictions compared to the input kinematics.  To mimic the procedure used in orbit modeling of real data, we do not use regularization in this step (see discussion in Sec~2.2). 
The metric, hereafter referred to as the log-likelihood, is given by

\begin{equation}\label{eqn:logl}
-2 \ln \mathcal{L}(\vect{\mu}) = \chi^2_\textrm{kin}(\vect{\mu}) = \sum_j^{\nbin}\sum_i^{N_\textrm{GH}} \frac{(h_{ij}(\vect{\mu}) - h_{ij,\rm{input}})^2}{\Delta h_{ij,\rm{input}}^2} \,,
\end{equation}

where $h_{ij}$ is the $i$-th Gauss-Hermite moments of the stellar
LOSVDs in the $j$-th spatial bin, $\Delta h_{ij}$ is the associated measurement uncertainty, and $\vect{\mu}$ 
is the set of six parameters describing the galaxy's potential.
For the template galaxy NGC 2693 used in this work, 
Gemini GMOS integral-field spectroscopy (IFS) yielded 60 spatial bins, each with 8 Gauss-Hermite moments, for the central $5''\times 7''$ region of the galaxy, and
McDonald Mitchell IFS yielded 29 bins, each with 6 moments, for the outer part of the galaxy \citep{pilawaetal2022};
together, the total number of kinematic constraints is $\nkin = 654$.

Adopting the parameter inference procedure used in our prior work \citep{quennevilleetal2022, pilawaetal2022, liepoldetal2023},  we first construct an interpolated log-likelihood surface from the discrete set of evaluated models using Gaussian process regression \citep{rasmussenwilliams2006} with a Mat\'ern covariance kernel with $\nu=3/2$.  The resulting Gaussian process mean function is used as a smooth surrogate function for the true log-likelihood surface.  The dynamic nested sampler \texttt{dynesty} \citep{speagle2020} is then used to sample this smooth log-likelihood function and to produce a posterior assuming uniform priors on each parameter.

A typical outcome of our recovery tests is shown in Figure~\ref{fig:N2693_sample_6d}, where the 1D and 2D marginalized posteriors for the six model parameters in realization $r_5$ of model $\textrm{G}_4$ are plotted.
The true input values of the model parameters (orange circles) are well recovered, lying within the 95\% confidence interval of the best-fit model for all six parameters.  

The results for all 25 recovery tests are summarized in Figure~\ref{fig:N2693_Triaxial_Recovery}.
\cpm{The input parameter values are shown in the legend of each panel and in Table~\ref{tab:mock_table}.
Overall, these results show an excellent recovery of the input parameters for (i) individual mock realizations (small dots), (ii) when averaging over five realizations for each galaxy model ${\rm G}_i$ (big circles), and (iii) when averaging over all 25 tests (big square).  
In particular, the two key mass parameters -- black hole mass and stellar mass-to-light ratios -- are consistently recovered within the 68\% confidence interval.  For other parameters, while there are some variations across different realizations and galaxy models, the gray squares indicate that the 
\textit{average} bias is consistent with 0 for all parameters.  
}

In addition to recoveries of galaxy model parameters, we can 
also assess the ability of the \TriOS\ code to 
recover intrinsic properties of the simulated galaxies. 
One such key property is the velocity anisotropy, typically parameterized by $\beta \equiv 1-\sigma_t^2/\sigma_r^2$, where $\sigma_r$ is the radial velocity dispersion, and  $\sigma_t$ is the tangential velocity dispersion with $\sigma_t^2 \equiv (\sigma_\theta^2 + \sigma_\phi^2)/2 $.  The value $\beta = 0$ indicates velocity isotropy, while a negative (positive) $\beta$ indicates tangential (radial) anisotropy.

\begin{figure}[htp!]
  \centering
 \includegraphics[width=0.47\textwidth]{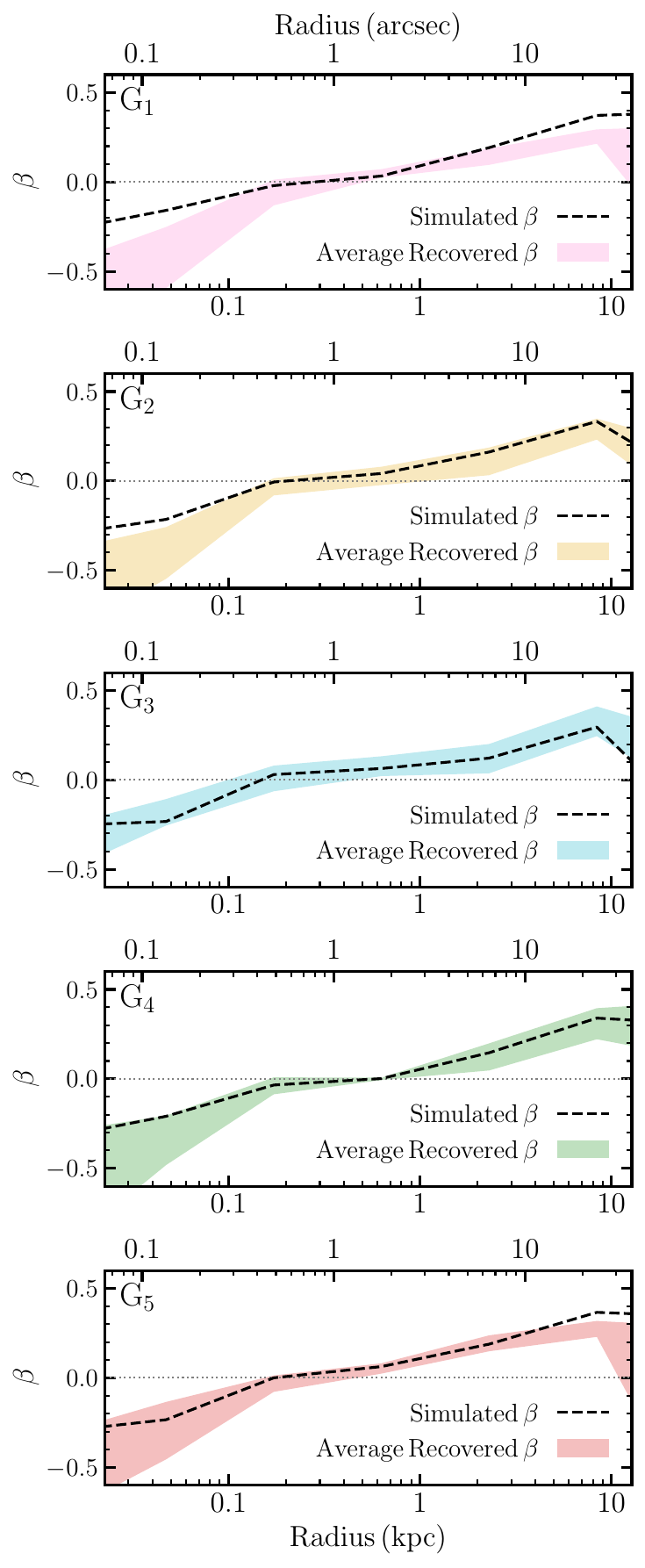}
    \caption{ Profile of the velocity anisotropy parameter, $\beta \equiv 1-\sigma_t^2/\sigma_r^2$, for the five simulated galaxies $G_1$ to $G_5$ (top to bottom; black dashed curves).  At each radius, the colored band indicates the standard deviation of $\beta$ from the recovery tests of the five realizations for the given galaxy model.
 }
    \label{fig:beta_recovery}
\end{figure}

Figure~\ref{fig:beta_recovery} shows the velocity anisotropy of the five simulated galaxies (black dashed curve in each panel) is preferentially tangential $\left(\beta < 0\right)$ near the centers of the galaxies and preferentially radial $\left(\beta > 0\right)$ beyond $\sim 1$ kpc. 
The $\beta$ profiles from the five lowest $\chi^2$ models (one for each $r_i$) in our recovery tests are shown as a solid color band in each panel, with the band corresponding to the mean and standard deviation in $\beta$ at a given radius. We include $\beta = 0$ as the dotted gray line to guide the eye. 
Our tests recover the simulated $\beta$ profiles quite well, with the largest discrepancies near the innermost and outermost portions of the galaxy. This trend in the innermost region has been noticed in anisotropy recovery studies before, but has been attributed to a number of different factors, including the relative paucity of orbits/mass in this innermost region \citep{Breddels2013, Kowalczyk2017}.

One feature in Figure~\ref{fig:N2693_Triaxial_Recovery} is the tendency towards over-estimation of the dark matter mass parameter \mdm\ in three of five galaxy models (G1, G4, G5) at the ${\sim}1\sigma$ level. 
While this trend could be simply due to small number statistics,
there are factors that could make the halo parameter less well constrained than other parameters.
In general, the constraint on the halo is driven by the outermost portion of the kinematic data since this is the region in which the enclosed dark matter mass starts to become comparable to that of the stars.  
Due to the rapidly decreasing surface brightness in the outer regions of galaxies, we must co-add stellar spectra over a much larger area of the sky to achieve a reasonable signal-to-noise ratio.  Furthermore, our outer data were obtained from the wide-field 
Mitchell IFS on a 2.7-meter telescope, while the inner data were from the high-resolution GMOS on a 8.1-meter telescope.
The larger uncertainties in the outermost kinematic data points thus allow for larger perturbations when we generate the simulated galaxy kinematics. \cpm{In addition, our recipe for computing the simulated Gauss-Hermite moments thus far does not take into account the correlations in the {\it uncertainties} in the moments \citep{houghtonetal2006}.}
These factors together can potentially bias the halo parameter more than the other parameters in our model.

\section{Model Complexity and Impact of Penalty Terms}

As discussed in Sec.~1, in the framework of classical statistic modeling, increasing the complexity of a model generally results in a better goodness-of-fit statistic. If the complexity is variable across models, the apparent goodness-of-fit may be biased by the variable complexity.
In this section, we examine this issue by adding a term to our log-likelihood function of Equation~(\ref{eqn:logl}) in the form of
\begin{equation}\label{eq:mod_like}
  -2\ln \mathcal{L}(\vect{\mu}) \rightarrow -2\ln \mathcal{L}(\vect{\mu}) + \kappa(\vect{\mu}) \,,
\end{equation}
where $\kappa$ is a ``penalty'' term used to capture a model's complexity. Various forms of $\kappa$ have been proposed in the literature (see Sec.~1) 
to penalize the models in the parameter space with more complexity and alleviate any resultant biases in the inferred model parameters.

Intuitively, $\kappa$ can be interpreted as serving a similar role as the Bayesian prior probability distribution. 
Just as the prior changes the shape of the posterior probability distribution given some previous knowledge of the parameter space, the shape of the penalty term
$\kappa(\vect{\mu})$
adjusts the goodness-of-fit landscape in our study,
$\chi^2_\text{kin}(\vect{\mu})$,
to correct for variability in model complexity as a function of the parameter space location $\vect{\mu}$.
In regions in which the model complexity $\kappa$ is high, the goodness-of-fit $\chi^2_\text{kin}$ should tend to be lower, and the relative trade-off of these two quantities yields the final inferred locations by subtly displacing the minimum of the modified log-likelihood landscape.

\cpm{Here we consider the complexity measure proposed in \citet{ye1998} and adopted in \citet{lipkaetal2021}. Those works compute a generalized DOF that is related to the ability of the model to respond to small perturbations in the model constraints. Intuitively, a model with a higher complexity (or flexibility) will be more responsive to those small perturbations, while a simple (or inflexible) model will be less responsive.
In our application, the constraints are given by the observed stellar kinematic moments and the goodness of fit is described by the log-likelihood. To measure the flexibility of a given model, we perturb the stellar kinematic predictions from that model a refit the model given those perturbed kinematics. The complexity (or flexibility) of the model is related to the amount of improvement in the log-likelihood after this re-fitting. For flexible models the improvement is large while for inflexible models the improvement is smaller.
This scheme empirically connects variations in the data directly to variations in the goodness of fit on a model-by-model basis without making significant assumptions about the form of the modeling.

\begin{figure*}[htp]
  \centering
  \includegraphics[width=5in]{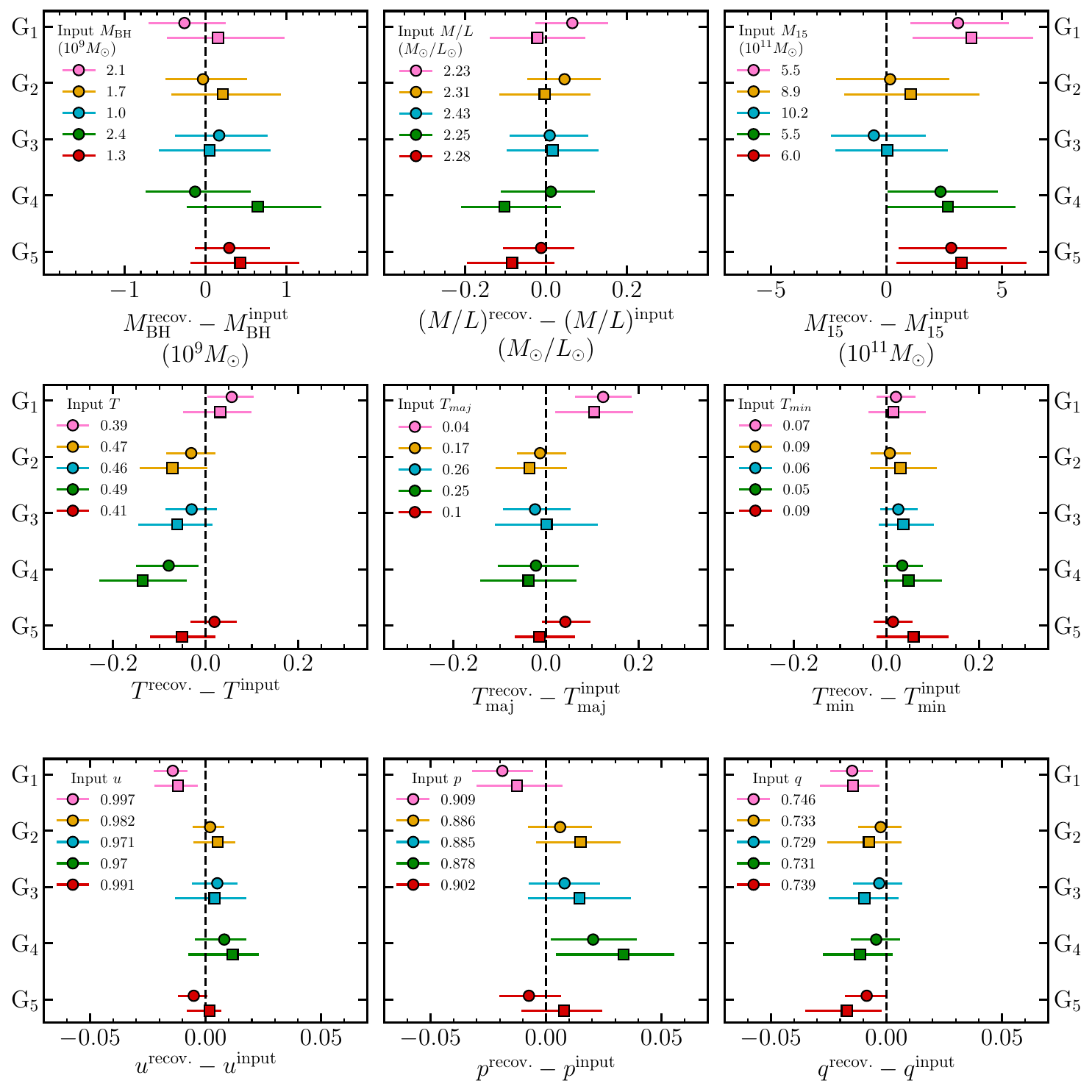}
    \caption{ Effect of adding a penalty term to the log-likelihood function (see Equation 2) in the recovery tests. The filled circles are identical to those in Figure~2 and
    denote the mean difference between the recovered and true parameter values of the five noise realizations for each galaxy model $\textrm{G}_i$.  The corresponding result after including a penalty term is shown as squares. 
    A penalty term leads to no improvement in parameter recovery.
    }
\label{fig:N2693_Triaxial_meff_comparison}
\end{figure*}

The specific steps used to compute the penalty term are as follows.}
We first fit a set of galaxy kinematics using a Schwarzschild model with parameters $\vect{\mu}$. We then perturb this model's kinematic moment predictions using Gaussian-distributed random values whose amplitudes are equal to the original measurement errors. 
We generate 20 realizations of these perturbations. For each realization, a $\chi^2$-like term $\chi^2_\textrm{prior,\textit{i}}$ is computed, which describes the difference between the perturbed and unperturbed moments. This quantity only depends on the number of kinematic moments that  have been perturbed, with $\langle \chi^2_\textrm{prior, \textit{i}} \rangle = \nkin$. We then re-optimize the orbital weights in the Schwarzschild model to fit these perturbed kinematics, resulting in a goodness-of-fit $\chi^2_\textrm{posterior, \textit{i}}(\vect{\mu})$. The difference in these two quantities is taken to be a measure of the complexity with {$\chi^2_\text{prior} - \chi^2_\text{posterior}(\vect{\mu}) \equiv \meff(\vect{\mu})$}, appropriately averaged across the 20 realizations. For a given set of galaxy model parameters, this procedure is identical to generating a simulated galaxy with those parameters, then fitting the simulated data with the same set of parameters. This is repeated for each of the ${\sim}3000$ models in our sample of orbit models used for the analyses in previous section. 

The recovered parameters for the five simulated galaxies using the original likelihood $\mathcal{L}$ (circles; same as in Fig.~\ref{fig:N2693_Triaxial_Recovery}) versus the penalized likelihood (squares) are shown in Figure~\ref{fig:N2693_Triaxial_meff_comparison}. 
Each symbol represents the mean value (relative to the true value) over the five noise realizations, and the uncertainty is taken to be the median uncertainty of the realizations.
Overall, there are only small differences in the resulting best-fit parameters and confidence intervals, and the average recovered values and median uncertainties are all consistent with one another at the $68\%$ confidence level. 
The surface of our penalty term as a function of galaxy parameters is generally quite flat and shows no consistent structure. 
In some cases in Figure~\ref{fig:N2693_Triaxial_meff_comparison}, adding the penalty term in fact moves the recovered median value away from the true value. 
This first exploration of model complexity for triaxial models therefore indicates that both the mass and shape parameters of a simulated triaxial galaxy are well recovered using our likelihood function without a penalty term.  

When axisymmetry is imposed in orbit models, \citet{lipkaetal2021} find their penalty term to be essentially monotonically increasing with increasing galaxy inclination angle $i$.  
The source of this dependence was attributed to the orbital structures of an axisymmetric potential as follows. A fair phase-space sampling of an axisymmetric galaxy includes both a prograde and a retrograde copy of every orbit. In the edge-on axisymmetric limit, these two orbits have opposite signs in the line-of-sight velocities and can contribute maximally to the model LOSVDs. In the face-on limit, on the other hand, the velocities of these two orbits are perpendicular to the line of sight, and the prograde and retrograde orbits contribute essentially identically to the resultant LOSVD. The inclination-dependent degeneracy between the prograde and retrograde orbits yields an effectively smaller number of distinct orbits for oblique and maximally face-on inclinations than edge-on inclinations, resulting in models with edge-on inclinations having more flexibility to produce good fits to the input data.

Triaxial potentials admit a larger suite of orbital types with more complicated structures and less symmetry than axisymmetric potentials. The inclination angle for an axisymmetric galaxy is replaced by three angles, or equivalently, three shape parameters (e.g., \citealt{vandenboschetal2008, quennevilleetal2022}). While adding a penalty term improved the recovery of the inclination angle in axisymmetric mock models, we do not find such a term to be necessary for triaxial models.

\section{Conclusion}

We have presented a study on the accuracy and reliability of recovering known sets of input galaxy mass and shape parameters using the triaxial Schwarzschild orbit modeling code \texttt{TriOS}.
We generated 25 sets of simulated galaxy kinematics
and performed a full search in the 6D parameter space
using our likelihood-based inference scheme to derive best-fitting parameters and confidence intervals. 
Treating the goodness-of-fit as proportional to a Bayesian log-likelihood, we produced posterior probability distributions for each parameter using dynamic nested sampling. 
The computational cost of this study is 
similar to that of performing full-scale triaixal Schwarzshild orbit
modeling and parameter inference for 25 galaxies.

The outcome of this study indicates robust recovery of both the parameters defining the galaxy potential (Figures~\ref{fig:N2693_sample_6d} and \ref{fig:N2693_Triaxial_Recovery}) and internal orbital structures such as the stellar velocity anisotropy profiles (Figure~\ref{fig:beta_recovery}). 
In particular, the two
key mass parameters in the galaxy models -- black hole mass and the stellar mass-to-light ratio -- are always recovered within the 68\% confidence interval of the true values. 

We additionally estimated the statistical model complexity of the triaxial models and tested the impact of this penalty term on the resulting parameter estimation.   In contrast to recent tests of axisymmetric orbit models, we find that penalty terms quantifying the varying model-to-model complexity do not appreciably alter the shape of the likelihood landscape, nor do they significantly impact the location of the best-fitting models (Figure~\ref{fig:N2693_Triaxial_meff_comparison}). At best, the posteriors obtained with the inclusion of a penalty term reproduce the posteriors obtained without a penalty term; at worst, the addition of a penalty term artificially increases the sizes of the confidence intervals, weakening the overall statistical power of the models and parameter inference scheme. 

Taken together, our results suggest that
triaxial Schwarzschild orbit modeling with the \texttt{TriOS} code, when combined
with dense sampling of the 6D parameter space and  our parameter inference scheme,
can {\it simultaneously} recover known mass parameters (for black holes, stars, and dark matter) as well as triaxial galaxy shape parameters without the need of a penalty term to compensate for model complexities. 

\acknowledgments

The authors acknowledge support from NSF AST-2206307, the Heising-Simons Foundation, and the Miller Institute for Basic Research in Science.
This work used the Extreme Science and Engineering Discovery Environment (XSEDE) at the San Diego Supercomputing Center through allocation AST180041, which is supported by NSF grant ACI-1548562.

\software{Astropy \citep{astropy:2013, astropy:2018},
Dynesty \citep{speagle2020},
Matplotlib \citep{Hunter:2007},
NumPy \citep{harris2020array}.}

\newpage
\bibliography{mocks}

\end{document}